\documentclass[12pt]{article}

\usepackage{amssymb,amsmath,bm,bbm}
\usepackage{cite}
\usepackage{color}
\usepackage{slashed}
\usepackage{graphicx}

\allowdisplaybreaks[2]

\newcommand{\Fbar}{\overline{F}}

\newcommand{\vev}[1]{\left\langle #1\right\rangle}
\newcommand{\eps}{\epsilon}

\newcommand{\ord}[1]{\mathcal{O}\left( #1 \right)}
\newcommand{\gm}{\gamma^\mu}

\newcommand{\GeV}{\rm GeV}
\newcommand{\TeV}{\rm TeV}

\begin{document}
\begin{titlepage}
\vspace*{-1.0truecm}
\begin{flushright}
CERN-PH-TH/2011-117\\
FLAVOUR(267104)-ERC-1 \\
TUM-HEP-807/11
\end{flushright}

\vspace{0.4truecm}

\begin{center}
\boldmath

{\Large\textbf{FCNC Effects \\ \vspace{6pt}  in a Minimal Theory of Fermion Masses}}			

\unboldmath
\end{center}

\vspace{0.4truecm}

\begin{center}
{\bf Andrzej J.~Buras$^{a,b}$, Christophe Grojean$^{c,d}$,\\
Stefan Pokorski$^{e,c,b}$, Robert Ziegler$^{a,b}$
}
\vspace{0.4truecm}

{\footnotesize
 $^a${\sl Physik Department, Technische Universit\"at M\"unchen,
James-Franck-Stra{\ss}e, \\D-85748 Garching, Germany}\vspace{0.2truecm}

$^b${\sl TUM-IAS, Technische Universit\"at M\"unchen,  Lichtenbergstr.~2A,\\ D-85748 Garching, Germany \vspace{0.2truecm}}

$^c${\sl CERN Physics Department, Theory Division,\\ CH-1211 Geneva 23, Switzerland \vspace{0.2truecm}}

$^d${\sl Institut de Physique Th\'eorique, CEA/Saclay,\\ F-91191 Gif-sur-Yvette C\'edex, France \vspace{0.2truecm}}

$^e${\sl Institute of Theoretical Physics, Faculty of Physics, University of Warsaw, Ho\.za 69,\\ 00-681, Warsaw, Poland}

}
\end{center}

\vspace{0.4cm}
\begin{abstract}
\noindent
As a minimal theory of fermion masses we extend the SM by heavy vectorlike fermions, with flavor-anarchical Yukawa couplings, that mix with chiral fermions such that small SM Yukawa couplings arise from small mixing angles. This model can be regarded as an effective description of the fermionic sector of a large class of existing flavor models and thus might serve as a useful reference frame for a further understanding of flavor hierarchies in the SM. Already such a minimal framework gives rise to FCNC effects through exchange of massive SM bosons whose couplings to the light fermions get modified by the mixing. We derive general formulae for these corrections and discuss the bounds on the heavy fermion masses. Particularly stringent bounds, in a few TeV range, come from the corrections to the $Z$ couplings.

\end{abstract}
\end{titlepage}

\section{Introduction}
The Standard Model (SM) successfully explains the observed suppression of flavor changing neutral currents (FCNC) and CP violation effects. However, it says nothing about the origin of the hierarchies in the fermion masses and the CKM mixing angles. Several theoretical ideas have been proposed to explain the peculiar pattern of fermion masses and mixings, such as horizontal symmetries~\cite{HSymm}, fermion localization in extra dimensional models of Randall--Sundrum (RS) type~\cite{RSFlavor} or wave function renormalization (WFR)~\cite{DIU} by new strong interactions~\cite{NeSt}. In most of these models, the flavor hierarchies among SM fermion masses and mixings are generated directly or indirectly through their dynamical mixing with new heavy (vectorlike) fermions. SM Yukawa couplings then arise from mixing angles that are given as ratios of mass parameters. Apart from the heavy fermionic sector, typically also new bosonic degrees of freedom are introduced that give rise to new flavor changing neutral currents (FCNC) and CP violating effects at low energy (e.g. flavons, KK gluons or sfermions). Very good agreement of such processes with the SM predictions leave little room for new contributions and thus requires the additional dynamics to be sufficiently heavy. While the minimal fermionic sector consists of a certain number of fermions with the same SM quantum numbers as the light fermions, the bosonic sector is very model dependent and so are the predictions for the new contributions to flavor violating effects at low energy. 

We therefore find it useful to construct a ``minimal'' theory of fermion masses that is capable to explain the hierarchy of SM Yukawa couplings through mass hierarchies. For this purpose we extend the SM only by a heavy fermionic sector that mixes with chiral fermions, such that small Yukawa couplings arise from small mixing angles. This mechanism is generic for many complete flavor models such as Froggatt--Nielsen (FN) or RS models, and therefore our construction can serve as an effective description of the fermionic sector of a large number of these kind of models. Besides this aspect one can regard the model in a bottom-up approach as a minimal extension of the SM that contains only the essential pieces necessary to parameterize small couplings by small mass ratios.  Even though complete flavor models typically depart from the minimal structure, we think that this setup can serve as a useful reference frame for a further understanding of the origin of flavor hierarchies in the SM. 

Already this minimal framework gives in general rise to FCNC and CP violating processes beyond the SM predictions. Since the light SM fermions have admixture of heavy fermions which have explicit mass terms, it is clear that 1) SU(2) doublets mix with SU(2) singlets and 2) SM fermion masses are not aligned with Higgs boson couplings. This implies that the couplings of the light fermions both to the massive SM gauge bosons and the Higgs boson will receive corrections that in general are flavor non-diagonal, but suppressed by the ratio of electroweak scale and the heavy fermion mass. Therefore all massive SM bosons mediate new flavor changing processes at the tree-level. Since these effects are due to the same mixing that gives rise to SM fermion masses, they depend on the same (small) parameters that determine the SM Yukawa couplings. 

In order to construct a ``minimal'' model that merely contains the essential structure needed to parameterize Yukawa couplings in terms of mass hierarchies, we start by adding an unspecified number of vectorlike fermions to the SM and require that all dimension four operators that can be constructed are either absent or have couplings that are $\ord{1}$. This setup is general enough to effectively describe the fermionic sector of almost every possible flavor model. On the other hand, there are clearly too many unknown parameters which prevent the extraction of any useful information. We therefore try to reduce the number of parameters such that the resulting model can still reproduce SM Yukawas and in addition maximally suppresses flavor violating effects. In this way we can identify the minimal FCNC effects and allow the heavy fermions to be as light as possible. In a last step we restrict to the minimal number of heavy fermions needed to explain all SM masses. In this procedure we partially give up the original generality, but the resulting model provides a simple framework which allows to study the minimal phenomenological effects. It is then straightforward to include additional structures of complete flavor models.
\begin{figure}
\begin{center}
\includegraphics[width=0.45\textwidth]{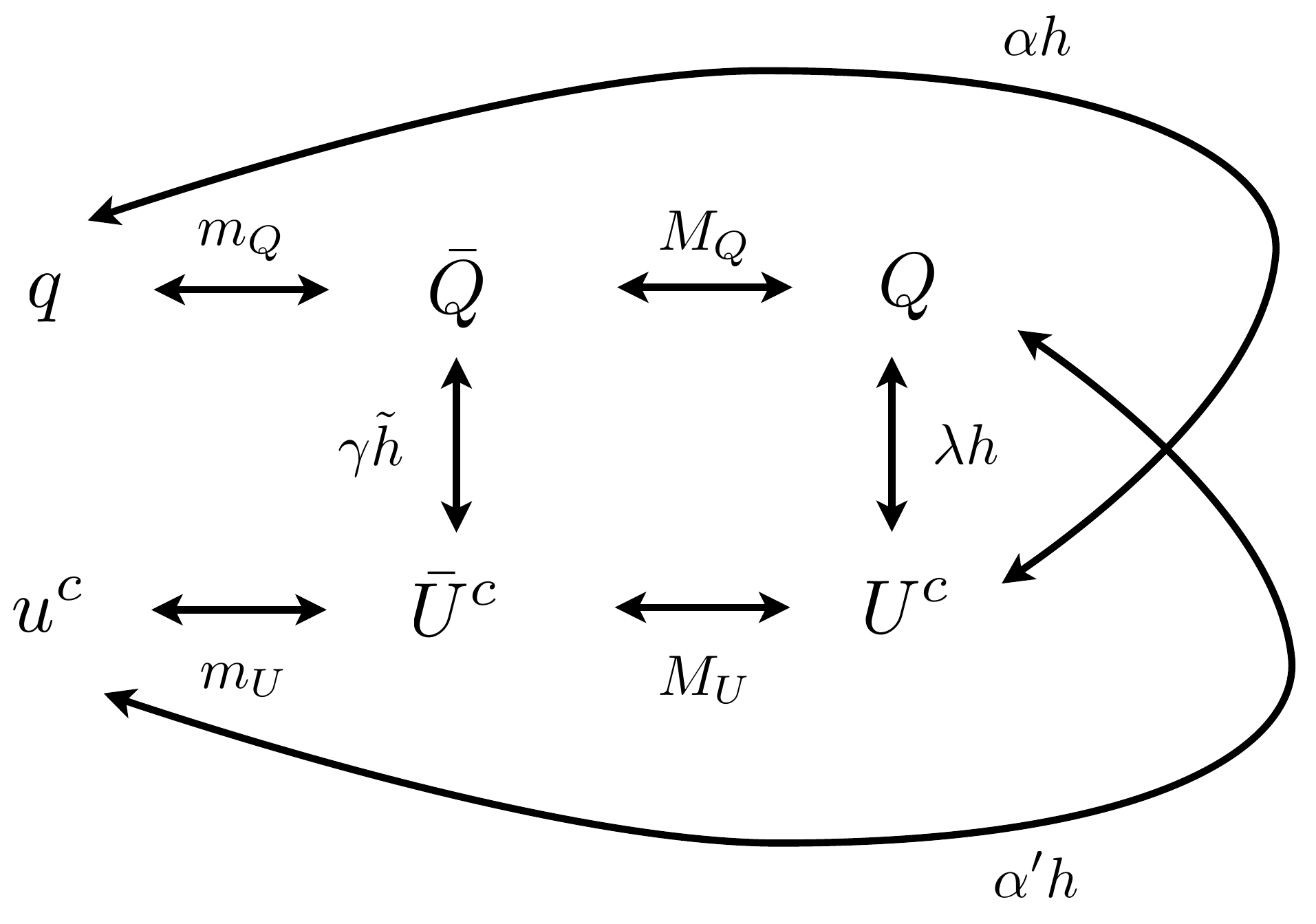}  
\caption{Graphical representation of the couplings in Lagrangian (1)}
\end{center}
\end{figure}
According to this procedure, which we outline here restricting to the up-sector and one family for simplicity, we start by adding vectorlike fermions\footnote{In this section we denote fermions with left-handed Weyl spinors.} $(Q + \overline{Q}, U^c + \overline{U^c})$ to the chiral field content of the SM $(q,u^c)$ and write the most general Lagrangian, that reads up to canonical kinetic terms
\begin{align}
\label{Lgen}
- {\cal L} & =  M_Q \overline{Q} Q + M_U \overline{U}^c U^c + m_U \overline{U}^c u^c  + m_Q \overline{Q} q \nonumber \\ & +  \lambda h Q U^c + \gamma \tilde{h} \overline{Q} \, \overline{U}^c + \alpha h U^c q + \alpha^\prime h Q u^c + {\rm h.c.},  
\end{align}
where $\tilde{h} \equiv i \sigma^2 h^*$, $(\lambda, \gamma, \alpha, \alpha^\prime)$ are $\ord{1}$ or vanishing and we take $m \lesssim M$ such that $u^c,q$ are predominantly light fermions. We do not include Yukawa couplings $y h q u^c$ since clearly we cannot get realistic masses for $y \sim \ord{1}$. Instead SM masses arise after electroweak symmetry breaking (EWSB) only through the mixing of light $(u^c,q)$ with heavy fermions $(U^c,Q)$ and are suppressed by small mixing angles $m/M$.  In order to see this explicitly, we integrate out the heavy fermions by their equations of motion\footnote{ We do not include kinetic terms and will recover their leading order effects by putting the solutions back into the heavy kinetic terms.}
\begin{subequations}
\begin{align}
Q & =   - \frac{1}{M_Q} \left[ m_Q q - \gamma \tilde{h} (h q) \left( \frac{\alpha}{M_U} - \lambda \frac{m_Q}{M_Q M_U} - \lambda \alpha \gamma \frac{h^\dagger h}{M_Q M_U^2} + \ord{1/M_Q^2 M_U^2} \right) \right] \label{Qbareom} \\
\overline{Q}  & = - \frac{\tilde{h}^*}{M_Q} \left[- \alpha^\prime + \lambda \left( \frac{m_U}{M_U} + \alpha^\prime \gamma \frac{h^\dagger h }{M_Q M_U} - \lambda \gamma \frac{m_U h^\dagger h }{M_Q M_U^2} + \ord{1/M_Q^2 M_U^2} \right) \right] u^c \\
U^c & =   - \frac{1}{M_U} \left[ m_U +  \alpha^\prime \gamma \frac{h^\dagger h }{M_Q} - \lambda \gamma m_U \frac{h^\dagger h}{M_Q M_U} - \lambda \gamma^2 \alpha^\prime \frac{(h^\dagger h)^2}{M_Q^2 M_U} + \ord{1/M_Q^2 M_U^2} \right] u^c  \\
\overline{U}^c  & = - \frac{1}{M_U} \left[ \alpha - \lambda \frac{m_Q}{M_Q} - \lambda \alpha \gamma \frac{h^\dagger h}{M_Q M_U} + \lambda^2 \gamma \frac{m_Q h^\dagger h}{M_Q^2 M_U} + \ord{1/M_Q^2 M_U^2}\right] h q.
\end{align}
\end{subequations}
The effective Lagrangian is then given by
\begin{align}
\label{effyuk}
{\cal L}_{\rm eff} & = \left[ \alpha \frac{m_U}{M_U} + \alpha^\prime \frac{m_Q}{M_Q} + \alpha \alpha^\prime \gamma \frac{h^\dagger h }{M_Q M_U} - \lambda  \frac{m_Q m_U}{M_Q M_U} \right] h q u^c \nonumber \\
& - \left[ \alpha \lambda \gamma \frac{m_U h^\dagger h}{M_Q M_U^2} + \alpha^\prime \lambda \gamma \frac{m_Q h^\dagger h}{M_U M_Q^2} + \alpha \alpha^\prime \lambda \gamma^2 \frac{(h^\dagger h)^2}{M_Q^2 M_U^2} - \lambda^2 \gamma \frac{m_U m_Q h^\dagger h}{M_Q^2 M_U^2}\right] h q u^c,
\end{align}
up to terms $\ord{1/M_Q^3 M_U^2,1/M_Q^2 M_U^3}$ and a part from the kinetic terms that we will discuss in a moment. Since we require that all couplings that do not vanish are $\ord{1}$ and flavor-anarchic, we have to set $\alpha = \alpha^\prime = 0$, since otherwise SM masses would be given dominantly by the first 2 terms in Eq.~(\ref{effyuk}), and we cannot fit fermion masses and mixings with such a structure. We also want to require that FCNCs are as much suppressed as possible. One immediate contribution to FCNCs arises from the last term in Eq.~(\ref{effyuk}) which gives rise to ``Higgs-dependent'' Yukawa couplings and induces flavor non-diagonal Higgs couplings~\cite{BaNa,GiLe,Toharia,AgCo}.  In order to suppress this source of Higgs-mediated FCNCs we therefore set also $\gamma =0$. We are left with effective Yukawa couplings 
\begin{align}
y_{eff} = - \lambda  \frac{m_Q}{M_Q} \frac{m_U}{M_U}, 
\end{align}
and it is well known that one can fit fermion masses and mixings with this structure. 

Let us now discuss hows FCNCs arise in the effective theory. Since in the fundamental theory we mix SU(2) singlets with SU(2) doublets, it is clear that the couplings of light fields to the $Z$ boson receive corrections that in general are not flavor-diagonal. Moreover, light masses arise partially from explicit mass terms and therefore there is no reason why Higgs boson couplings should be aligned to the light masses. Therefore we expect deviations from the SM both in $Z$ and Higgs boson couplings, but since these effects are due to SU(2) breaking and must vanish when we decouple the heavy fermions, they are suppressed at least by $v^2/M^2$. In the effective theory, these effects originate from the kinetic terms of the heavy fermions. Inserting the solutions of the EOMs into the heavy kinetic terms we get a contribution to the effective Lagrangian
\begin{align}
\label{effkin}
\Delta{\cal L}_{eff} & = u^{c \dagger} \left( 1 + \frac{m_U^2}{M_U^2} \right)  i \slashed{D} u^c + q^\dagger \left( 1 + \frac{m_Q^2}{M_Q^2} \right) i \slashed{D}  q  \nonumber \\
& + \frac{\lambda^2}{M_Q^2} \frac{ m_U^2}{ M_U^2} u^{c \dagger} \tilde{h} \, i \slashed{D} \left( \tilde{h}^* u^c \right)  +  \frac{\lambda^2}{M_U^2} \frac{ m_Q^2}{M_Q^2} q^\dagger h^\dagger i \slashed{D} \left( h q \right),
\end{align}
where we have taken couplings and masses real for simplicity. The terms in the first line of Eq.~(\ref{effkin}) give just an overall rescaling factor for light fermions that will lead to higher order terms in $m/M$ in Lagrangian. The terms in the second line instead generate (in general flavor non-diagonal) corrections to the couplings of $u$ and $u^c$ to the $Z$ and Higgs boson (and $W$) which are of the form
\begin{align}
\label{deltaZ}
\delta g_{Zuu}  & = - \frac{1}{2} \lambda^2 \frac{v^2}{M_U^2} \frac{m_Q^2}{M_Q^2},  & \delta g_{Zu^cu^c}  & = - \frac{1}{2} \lambda^2 \frac{v^2}{M_Q^2}\frac{m_U^2}{M_U^2},
\end{align}
\begin{align}
\label{deltaH}
\delta y_{H u u^c } & = - \frac{3}{2}  y_{\rm eff } \lambda^2 \left( \frac{v^2}{M_U^2} \frac{m_Q^2}{M_Q^2} + \frac{v^2}{M_Q^2}\frac{m_U^2}{M_U^2} \right),
\end{align}
as well as corrections to light fermion masses 
\begin{align}
\delta m_u  =  - \frac{1}{2}  y_{\rm eff } \lambda^2 \left( \frac{v^2}{M_U^2} \frac{m_Q^2}{M_Q^2} + \frac{v^2}{M_Q^2}\frac{m_U^2}{M_U^2} \right).  
\end{align}
Note that the effective Higgs couplings are no longer aligned to the light masses because they receive corrections that have different numerical factors. Therefore we have in general both $Z$ and Higgs mediated FCNC effects. 

Let us now shortly comment on the consequences for FNCNs when we allow for $\gamma \ne 0$. In this case the corrections to the $Z$ couplings are not modified in leading order, while from the last term in Eq.~(\ref{effyuk}) we get contributions to the Higgs boson couplings that are suppressed only by a factor $m^2/M^2$ instead of $m^4/M^4$ for the case $\gamma=0$. 
\begin{align}
\label{deltaHgamma}
\delta y_{H u u^c } & = - 3  y_{\rm eff } \lambda \gamma \frac{v^2}{M_Q M_U}. 
\end{align}
In Table~\ref{tab:summary} we collect the parametric suppression factors in the corrections to $Z$ and Higgs couplings for different choices of parameters. These results agree and extend the results obtained in Ref.~\cite{AgCo}.

\begin{table}[ht]
\begin{center}
$
\begin{array}{ccc}
\hline
\hline
\textrm{Couplings} \ne 0 & \hspace{1.5cm} \delta g_{Zf_i\bar{f}_j} ~(Z\textrm{~FCNC})&  \hspace{1cm}  \delta y_{Hf_i\bar{f}_j} ~(\textrm{Higgs FCNC})\\
\hline
{\vrule height 20pt depth 10pt width 0pt}
\alpha, \alpha^\prime, \gamma,\lambda & \displaystyle \frac{v^2}{M^2} &  \displaystyle \frac{v^2}{M^2} \\[.3cm]
\gamma, \lambda  & \displaystyle \frac{m_i m_j}{M^2} \frac{v^2}{M^2} & \displaystyle \frac{m_i m_j}{M^2} \frac{v^2}{M^2} \\[.3cm]
\lambda & \displaystyle \frac{m_i m_j}{M^2} \frac{v^2}{M^2} & \displaystyle \frac{m^2_i m^2_j}{M^4} \frac{v^2}{M^2}\\[.3cm]
\hline
\hline
\end{array}
$
\end{center}
\label{tab:summary}
\vspace{-.5cm}
\caption{\small Parametric dependence of the flavor-violating corrections to the $Z$ and the Higgs couplings in the presence of the Yukawa interactions in the vector-like sector, $\gamma$ and $\lambda$, and between the vector-like and the chiral sector, $\alpha$ and $\alpha'$.}
\end{table}

To summarize, for a minimal model of fermion masses we consider the schematic Lagrangian 
\begin{align}
\label{Lmin}
{\cal L} \sim m f \Fbar + M F \Fbar + \lambda h F F,
\end{align}
where $f=(q,u^c,d^c)$, $F=(Q,U^c,D^c)$ and the appropriate gauge structure is understood. Since $m$ is a rank three matrix the minimal number of heavy fermions needed is three for each $Q,U,D$ sector. We can further suppress flavor violating effects and reduce the number of parameters if we take $m$ to be approximately diagonal in the same basis as $M$. Then $m$ and $M$ leave a global $U(1)^3$ flavor symmetry unbroken which is violated only by the heavy Yukawa couplings $\lambda$, so that all flavor violating effects in the light sector are doubly suppressed by small mixing angles $m/M$.  This setup then defines our minimal model, containing only the necessary ingredients needed to explain hierarchical Yukawa couplings through mass hierarchies.

The rest of this paper is organized as follows. In Section 2 we discuss the minimal model in more detail and derive the corrections to the gauge-fermion and Higgs-fermion couplings that arise from the presence of the heavy fermions. In Section 3 we use these results to obtain approximate lower bounds on the masses of heavy fermions. It turns out that strong bounds on these masses arise from the corrections to the $Z$ couplings and are in a few TeV range. A detailed phenomenological analysis is in progress and will appear elsewhere \cite{BGPZ2}. In Section 4 we discuss the connection of the minimal model to existing flavor models and in Section 5 we envisage some additional structures that can be present in the heavy fermionic sector. In particular a model with a unitary Yukawa matrix in the heavy quark sector appears interesting from the point of view of a further reduction of fundamental parameters. We finally conclude in Section 6. 

\section{A Minimal Model}
We now construct an explicit model according to the philosophy described above where we restrict our discussion to the quark sector for simplicity. The chiral field content is given by three families of quarks\footnote{In this and the next section we use 4-component Dirac spinors.}
\begin{equation}
u_{Ri},d_{Ri},q_{Li} = \binom{u_{Li}}{d_{Li}} \hspace{1cm} i=1,2,3
\end{equation}
and we add for each chiral fermion a vectorlike pair of heavy quarks
\begin{equation}
U_{Ri},U_{Li},D_{Ri},D_{Li},Q_{Ri}= \binom{U_{Ri}^Q}{D_{Ri}^Q},Q_{Li}= \binom{U_{Li}^Q}{D_{Li}^Q} \hspace{1cm}  i=1,2,3. 
\end{equation}
In addition to (canonical) kinetic terms the Lagrangian is of the form 
\begin{align}
\label{L3}
- {\cal L} & = \tilde{h} \lambda^U_{ij} \bar{Q}_{Li} U_{Rj} + h \lambda^D_{ij} \bar{Q}_{Li} D_{Rj}  + M^U_{ij}  \bar{U}_{Li} U_{Rj} + M^D_{ij}\bar{D}_{Li}  D_{Rj}  + M^Q_{ij} \bar{Q}_{Ri} Q_{Lj} \nonumber \\
& + m^U_{ij}  \bar{U}_{Li} u_{Rj} + m^D_{ij}  \bar{D}_{Li} d_{Rj} + m^Q_{ij}  \bar{Q}_{Ri} q_{Lj} + {\rm h.c.}
\end{align}
with complex $3 \times 3$ matrices $\lambda^{U,D}, M^{U,D,Q}, m^{U,D,Q}$ and $\tilde{h} \equiv i \sigma^2 h^*$. We neglect other possible couplings and assume that their absence is justified by symmetries or other dynamical reasons. Moreover we assume that $m$ and $M$ in each sector are approximately diagonal in the same basis, for example as a consequence of an approximate degeneracy of the heavy fermions in each sector. At the end of this section we will comment about the consequences of relaxing these assumptions. 

Going into the basis where $m$ and $M$ are diagonal, we can absorb possible phases in the diagonal matrices with field redefinitions, so we arrive at the Lagrangian
\begin{align}
\label{Lorigin}
- {\cal L} & = \tilde{h} \lambda^U_{ij} \bar{Q}_{Li} U_{Rj} + h \lambda^D_{ij} \bar{Q}_{Li} D_{Rj}  + M^U_{i}  \bar{U}_{Li} U_{Ri} + M^D_{i}\bar{D}_{Li}  D_{Ri}  + M^Q_{i} \bar{Q}_{Ri} Q_{Li} \nonumber \\
& + m^U_{i}  \bar{U}_{Li} u_{Ri} + m^D_{i}  \bar{D}_{Li} d_{Ri} + m^Q_{i}  \bar{Q}_{Ri} q_{Li} + {\rm h.c.}
\end{align}
where $M_i$ and $m_i$ are diagonal matrices with positive entries. Instead of $m_i$ we will use 
\begin{align}
\eps_i^{Q,U,D} \equiv \frac{m_i^{Q,U,D}}{M_i^{Q,U,D}} 
\end{align}
which are also real and positive. Counting parameters, we have 18 real parameters from the masses and 18 real parameters plus 18 phases from the heavy Yukawas $\lambda^{U,D}$. We still have the freedom to do phase redefinitions for $U_R,D_R,Q_L$ so we get in total 36 real parameters and 10 phases. However, 18 of these real parameters are the heavy Yukawa couplings $\lambda^{U,D}$ which we require to be anarchical  $\ord{1}$ numbers. For the phenomenological analysis to be performed in Section~\ref{pheno} we will denote the overall strength of these couplings by 2 single parameters $\lambda^U_*,\lambda^D_*$ and neglect the flavor dependence. Moreover we will take the 9 heavy fermion masses to be universal in each sector, reducing their number to three $(M_Q,M_U,M_D)$. Finally out of the 9 $\eps_i$ parameters 8 are determined by masses and mixings (there is one prediction), leaving only $\eps_3^Q$ free. Thus we will describe the model by six real parameters $\lambda_*^U,\lambda_*^D, M_Q,M_U,M_D, \eps^Q_3$, in addition to the SM ones. 

Our next goal is to find the corrections to the SM couplings originating from the presence of heavy vectorlike fermions.  Rather general formulae of this type have been derived in Refs.~\cite{Aguila,BDG} assuming the presence of direct Higgs couplings to SM fermions. Our formulae below extend these considerations to the cases in which such couplings are absent and Yukawa couplings are only generated through mixing with heavy fermions. A detailed derivation of these results will be presented in Ref.~\cite{BGPZ2}.

In order to derive the low-energy effective Lagrangian, we first go to the mass basis in the limit of vanishing Higgs VEV $v \rightarrow 0$ and integrate out the heavy states using their equations of motion. Then we include EWSB with $v \approx 174$ GeV and finally redefine light fields to get canonical kinetic terms. This redefinition brings electromagnetic currents back to the standard form, while charged and neutral currents get new contributions due to the mixing of SU(2) doublets and singlets. In addition one finds (multi-) Higgs couplings that are not aligned to mass terms because of the presence of explicit mass terms in the original theory. The resulting low-energy Lagrangian up to $\ord{v^4/M^4}$ corrections and multi-Higgs couplings is given by
\begin{subequations}
\begin{align}
- {\cal L}_{\rm eff} & \supset \frac{g}{\sqrt2} \left( W^+_\mu j^{\mu -}_{\rm charged} + {\rm h.c.} \right) +  \frac{g}{2 c_{\rm w}} Z_\mu j^\mu_{\rm neutral} \nonumber \\
& + \overline{u}_{Li}m^U_{ij} u_{Rj} +  \overline{d}_{Li} m^D_{ij} d_{Rj}  + \frac{H}{\sqrt2} \left( \overline{u}_{Li} y^U_{ij} u_{Rj} +  \overline{d}_{Li} y^D_{ij} d_{Rj} \right) + {\rm h.c.}   
\end{align}
\begin{align}
\label{effm}
m^X_{ij}  & = v {\bar \eps}^Q_i {\bar \eps}^X_j \lambda^X_{ij}  -\frac{v}{2} \left( A^X_L \right)_{ik}  {\bar \eps}^Q_k {\bar \eps}^X_j \lambda^X_{kj} -\frac{v}{2} \left( A^X_R \right)_{kj}  {\bar \eps}^Q_i {\bar \eps}^X_k \lambda^X_{ik} \\
\label{effy}
y^X_{ij} & = \frac{m^{X}_{i j}}{v} - \left( A^X_L \right)_{ik}  {\bar \eps}^Q_k {\bar \eps}^X_j \lambda^X_{kj} -\left( A^X_R \right)_{kj} {\bar \eps}^Q_i {\bar \eps}^X_k \lambda^X_{ik}
\end{align}
\begin{align}
j^{\mu -}_{\rm charged} & = \overline{u}_{Li} \left[ \delta_{ij} - \frac{1}{2} \left(A^U_L\right)_{ij} - \frac{1}{2} \left(A^D_L\right)_{ij} \right] \gm d_{Lj} + \overline{u}_{Ri}  \left( A^{UD}_R\right)_{ij} \gm d_{Rj}  \\
j^\mu_{\rm neutral} & = \overline{u}_{Li} \left[ \delta_{ij} - \left(A^U_L\right)_{ij}  \right] \gm u_{Lj} + \overline{u}_{Ri}  \left( A^U_R\right)_{ij}  \gm u_{Rj}   \nonumber \\
& - \overline{d}_{Li} \left[ \delta_{ij} - \left(A^D_L\right)_{ij}  \right] \gm d_{Lj} -  \overline{d}_{Ri} \left( A^D_R\right)_{ij} \gm d_{Rj} - 2 s_{\rm w}^2 j^\mu_{\rm elmag} 
\end{align}
\begin{align}
\left( A^X_L \right)_{ij} & = \frac{v^2}{{\bar M}_k^X {\bar M}_k^X} {\bar \eps}^Q_i {\bar \eps}^Q_j  \lambda_{ik}^{X}  \lambda_{jk}^{X*}  &
\left( A^X_R \right)_{ij} & = \frac{v^2}{{\bar M}_k^Q {\bar M}_k^Q} {\bar \eps}^X_i {\bar \eps}^X_j  \lambda_{kj}^{X}  \lambda_{ki}^{X*} \label{ALXARX}\\
\left( A^{UD}_R\right)_{ij} & = \frac{v^2}{{\bar M}_k^Q {\bar M}_k^Q} {\bar \eps}^U_i {\bar \eps}^D_j  \lambda_{kj}^{D}  \lambda_{ki}^{U*} \\
\bar{\eps}_i^X & = \frac{\eps_i^X}{\sqrt{1+ \eps^X_i \eps^X_i}} & \bar{M}_k^X & = M_k^X (1 + \eps^X_k \eps^X_k) \label{Qeqn}
\end{align}
\end{subequations}
where $H$ is the Higgs boson, $X=U,D$ and in Eq.~(\ref{Qeqn}) $X=Q,U,D$. Note that the new contributions to Higgs and massive gauge boson couplings are doubly suppressed by small mixing angles $\eps_i$, which resembles the structure of flavor suppression in WFR and RS models and is a direct consequence of requiring the absence of light-heavy Yukawa couplings, i.e. taking $\alpha = \alpha' = 0$ in (\ref{Lgen}). 

Finally let us shortly comment on the deviations from the above formulae when we depart from our minimal framework. First we have neglected operators of the form $\gamma h \bar{Q}_R D_L$. They would give rise to additional contributions to flavor-violating fermion-Higgs couplings that would be suppressed only by $\eps^2$ instead of $\eps^4$, cf. Eqns~(\ref{deltaH}),(\ref{deltaHgamma}). However, as we will see in the next section, the resulting contributions to FCNCs are not much larger than the ones from flavor violating $Z$ couplings, so that the bounds on heavy fermion masses do not dramatically change when we allow for $\gamma \ne 0 $. Second we have assumed that $m$ and $M$ are diagonal in the same basis. The presence of off-diagonal entries of $m$ in the basis where $M$ is diagonal would merely induce corrections to the above formulae that are proportional to ratios of off-diagonal over diagonal entries. Provided that these ratios are not large, our expressions are approximately valid also in the case that the alignment of $m$ and $M$ is not exact.  

\section{Phenomenology}
\label{pheno}
We now make a rough estimate of the FCNC effects induced by the effective Lagrangian up to $\ord{1}$ coefficients and leave a more detailed analysis to a future publication~\cite{BGPZ2}. Here we are only interested in obtaining a lower bound for the masses of the heavy fermions $M^{Q,U,D}$. 

We first note that the leading order expression for SM masses
\begin{align}
\label{masses}
m^X_{ij}  \approx v \eps^Q_i \eps^X_j \lambda^X_{ij} \hspace{2cm} (X=U,D)  
\end{align}
reproduces the Yukawa structure in FN models, and it is well known that one can fit all SM masses and mixings when $\lambda_{ij}$ are $\ord{1}$ couplings and the $\eps$'s are certain powers of a small order parameter, e.g. the Cabibbo angle. In particular, since the top Yukawa coupling is large and CKM matrix elements are given by $V_{ij} \sim \eps_i^Q/\eps_j^Q$ for $i \le j$, we take \begin{align}
\eps^Q_1 & \sim \eps^3 \eps^Q_3 & \eps^Q_2 & \sim \eps^2 \eps^Q_3,                                                                                                                                                                                                                                                                                                                                                                                                                                                                                                                                                                                                                                                                                                                                                                                             \end{align}
with $\eps \approx 0.23$ keeping $\eps^Q_3$ as a free parameter that is $\lesssim 1$. The remaining six parameters $\eps^{U,D}_i$ are then determined through the six quark masses up to $\ord{1}$ couplings $\lambda_{ij}$. We write these masses as 
\begin{align}
\frac{m_i^X}{v} \sim \eps^Q_i \eps^X_i \lambda^X_*, 
\end{align}
where $\lambda^X_*$ represents the overall strength of the original couplings $\lambda_{ij}^X$ and is only included in order to keep track of the parametric dependence of our results on these couplings. Then we use the results of Ref.~\cite{Xing} to fit quark masses at 1 TeV by certain powers of $\eps$ (which are not uniquely determined since $\eps$ is not particularly small). We will take 
\begin{align}
\frac{m_u}{v} & \sim \eps^8 & \frac{m_c}{v} & \sim \eps^4 & \frac{m_t}{v} & \sim 1 \nonumber \\
\frac{m_d}{v} & \sim \eps^{7 \div 8} & \frac{m_s}{v}  & \sim \eps^{5 \div 6} &  \frac{m_b}{v} & \sim \eps^3, 
\end{align}
which in turn gives 
\begin{align}
\eps^U_1 & \sim \frac{\eps^5}{\eps^Q_3 \lambda_*^U} & \eps^U_2 & \sim \frac{\eps^2}{\eps^Q_3 \lambda_*^U} & \eps^U_3 & \sim \frac{1}{\eps^Q_3 \lambda_*^U} \nonumber \\
\eps^D_1 & \sim \frac{\eps^{4 \div 5}}{\eps^Q_3 \lambda_*^D} & \eps^D_2 & \sim \frac{\eps^{3 \div 4}}{\eps^Q_3 \lambda_*^D} & \eps^D_3 & \sim \frac{\eps^3}{\eps^Q_3 \lambda_*^D}. 
\end{align}
Since the constraints on right-handed charged currents (parameterized by $A^{UD}_R$) are rather weak, we concentrate in the following on Higgs and $Z$ couplings. We begin by rewriting the couplings to the Higgs scalar and the $Z$ boson in terms of SM masses in matrix notation
\begin{subequations}
\begin{align}
y^X = \frac{m^X}{v} - A^X_L \frac{m^X}{v} - \frac{m^X}{v} A^X_R 
\end{align}
\begin{align}
A^X_L  & = m^X B_L m^{X \dagger} & A^X_R  & = m^{X \dagger} B_R m^X  
\end{align}
\begin{align}
\label{Bs}
B_L & = \frac{1}{M_X^2}\begin{pmatrix} 
\frac{1}{\eps^X_1 \eps^X_1} & 0 & 0 \\ 
0 & \frac{1}{\eps^X_2 \eps^X_2} & 0 \\ 
0 & 0 & \frac{1}{\eps^X_3 \eps^X_3}  \\ 
\end{pmatrix} &
B_R & = \frac{1}{M_Q^2}\begin{pmatrix} 
\frac{1}{\eps^Q_1 \eps^Q_1} & 0 & 0 \\ 
0 & \frac{1}{\eps^Q_2 \eps^Q_2} & 0 \\ 
0 & 0 & \frac{1}{\eps^Q_3 \eps^Q_3}  \\ 
\end{pmatrix}, 
\end{align}
\end{subequations}
where we have neglected terms higher order in $\eps_i$ and assumed that heavy fermions are approximately degenerate in each sector
\begin{align}
M^{Q,U,D}_1 \approx M^{Q,U,D}_2 \approx  M^{Q,U,D}_3 \approx  M_{Q,U,D}. 
\end{align}
Going to the light mass basis defined by 
\begin{align}
V_L^{X \dagger} m^X V_R^X & = m^X_{diag},
\end{align}
we get for Higgs couplings in this mass basis
\begin{align}
\tilde{y}^X = \frac{m^X_{diag}}{v} - \tilde{A}^X_L \frac{m^X_{diag}}{v} - \frac{m^X_{diag}}{v} \tilde{A}^X_R.
\end{align}
The $Z$ couplings in the mass eigenstate basis read
\begin{align}
 \tilde{A}^X_L & = m^X_{diag} \tilde{B}_L m^X_{diag} & \tilde{A}^X_R & = m^X_{diag} \tilde{B}_R m^X_{diag} \\
\tilde{B}_L & = V_R^{X \dagger} B_L V_R^X & \tilde{B}_R & = V_L^{X \dagger} B_R V_L^X. \label{Btildes}
\end{align}
Up to $\ord{1}$ numbers determined by $\lambda_{ij}$ the dominant entries of the rotation matrices are $(V^X_L)_{ij} \sim \eps_i^Q/\eps_j^Q, \, (V^X_R)_{ij} \sim \eps_i^X/\eps_j^X$ for $i \le j$ and  $(V^X_L)_{ij} \sim \eps_j^Q/\eps_i^Q, \, (V^X_R)_{ij} \sim \eps_j^X/\eps_i^X$ for $i \ge j$. This gives 
\begin{align}
\left( \tilde{B}_L \right)_{ij} & \sim \frac{1}{M^2_X } \frac{1}{\eps^X_i \eps^X_j} & \left( \tilde{B}_R \right)_{ij} & \sim \frac{1}{M_Q^2} \frac{1}{\eps^Q_i \eps^Q_j}
\end{align}
and finally 
\begin{align}
\left( \tilde{A}^X_L \right)_{ij} & \sim  \frac{v^2}{M_X^2} \eps^Q_i \eps^Q_j (\lambda^X_*)^2 &
\left( \tilde{A}^X_R \right)_{ij} & \sim \frac{v^2}{M_Q^2} \eps^X_i \eps^X_j (\lambda^X_*)^2. \label{Atildes}
\end{align}
We are now ready to constrain the heavy masses using the bounds on FCNCs mediated by Higgs and $Z$. It is easy to see that Higgs mediated FCNC are negligible. An estimate of the coefficient of $(\bar s_R\, d_L)(\bar s_L d_R)/(1 \, \TeV)^2$ gives
\begin{align}
C_{LR} & \approx 12.5 \left(\frac{200 \, \GeV}{m_H}\right)^2 \frac{v^2}{M_D^2} \eps_1^Q \eps_2^Q  (\lambda_*^D)^4  \left( \frac{m_s m_d} {M_D^2} \eps_1^Q \eps^Q_2 + \frac{m_s^2} {M_Q^2} \eps_1^D \eps^D_2 \right) \nonumber \\
& \approx 12.5 \left(\frac{200 \, \GeV}{m_H}\right)^2 \frac{m_d m_s}{M_D^2} (\lambda_*^D)^2 \left( \frac{v^2} {M_D^2} \eps^{10} (\eps^Q_3)^4 (\lambda_*^D)^2 + \frac{m_s^2} {M_Q^2} \right),
\end{align}
where the quark masses are evaluated at the scale $m_H$. Thus $C_{LR}$ is 
of the order of $ 10^{-16} \left(  \TeV /M \right)^4 $ for $\eps^Q_3 \approx \lambda_*^D \approx 1$ and a Higgs mass of 115 GeV, to be compared with the bound from $\eps_K$ that constraints the imaginary part of this coefficient to be $\lesssim 3 \times 10^{-11}$~\cite{Bona, INP}. Similarly for $D^0-\bar D^0$ and $B^0_{d,s}-\bar B_{d,s}$ systems the resulting coefficients are far below the bounds in Refs.~\cite{Bona, INP}. 

Instead the strongest bounds on the heavy fermion masses arise from the presence of flavor off-diagonal $Z$ couplings that we write in the usual notation as 
\begin{align}
- {\cal L}_{\rm eff} \supset \frac{g}{c_W} Z_\mu \left( \delta g_L^{ds}  \overline{d}_L \gm  s_L +  \delta g_R^{ds} \overline{d}_R \gm  s_R +  \delta g_L^{uc}  \overline{u}_L \gm c_L +  \delta g_R^{uc}  \overline{u}_R \gm c_R \right)
\end{align}
with (note that $\delta g_{L,R} = - \frac{1}{2} \tilde{A}_{L,R}$) 
\begin{subequations}
\begin{align}
\delta g_L^{ds} & \sim \frac{1}{2} \frac{v^2}{M_D^2}  \eps^Q_1 \eps^Q_2 (\lambda_*^D)^2 \sim  1.5 \times 10^{-2} \, \eps^5 (\eps^Q_3)^2 (\lambda_*^D)^2 \left( \frac{1 \, \TeV}{M_D} \right)^2  \\ 
\delta g_R^{ds} & \sim \frac{1}{2} \frac{v^2}{M_Q^2}   \eps^D_1 \eps^D_2 (\lambda_*^D)^2  \sim  1.5 \times 10^{-2} \, \eps^{7 \div 9} (\eps^Q_3)^{-2} \left( \frac{1 \, \TeV}{M_Q} \right)^2 \\
\label{dguc}
\delta g_L^{uc} & \sim \frac{1}{2} \frac{v^2}{M_U^2}   \eps^Q_1 \eps^Q_2 (\lambda_*^U)^2 \sim  1.5 \times 10^{-2} \, \eps^5 (\eps^Q_3)^2 (\lambda_*^U)^2 \left( \frac{1 \, \TeV}{M_U} \right)^2 \\ 
\delta g_R^{uc} & \sim \frac{1}{2} \frac{v^2}{M_Q^2}   \eps^U_1 \eps^U_2 (\lambda_*^U)^2  \sim  1.5 \times 10^{-2} \, \eps^7 (\eps^Q_3)^{-2}  \left( \frac{1 \, \TeV}{M_Q} \right)^2. 
\end{align}
\end{subequations}
Constraints on $K_L \rightarrow \mu^+ \mu^-$ require $ |\delta g_{L,R}^{ds}| \le 6 \times 10^{-7}$ as we found using the upper bound in Ref.~\cite{IsUn}. This translates into 
\begin{align}
M_D & \gtrsim 4 \, \TeV \times \eps^Q_3 \lambda_*^D & M_Q \gtrsim  900 \, \GeV \times \frac{1}{\eps^Q_3}. 
\end{align}
Note that the bounds from $Z$ mediated $\Delta F = 2$ processes are weaker since the constraints on $\epsilon_K$ require $\delta g_{L,R}^{ds}$ to be only $\sim 10^{-5}$ at $M_Z$, giving bounds on $M_D$ and $M_Q$ that are weaker by a factor 4. Also constraints from other flavor observables are weaker, e.g. the corrections to $b \rightarrow s \gamma$ are below 1\% for $M \sim \ord{1 \, \TeV}$. What regards the bounds on $M_U$, large long distance contributions to the $D^0-\bar{D^0}$ mass  difference preclude any meaningful calculation of the lower bound on $M_U$. In fact values of $M_U$ as low  as few hundreds GeV cannot be excluded at present. 

We now compare the above bounds on heavy fermion masses coming from FCNC processes with the bounds obtained from electroweak precision data. As it has been recently emphasized in various studies performed in the context of RS models~\cite{Agashe:2003zs, Agashe:2006at, Carena:2007ua}, electroweak data also strongly constrain any deviation of the $Z$ coupling to the left-handed $b$ quark. In our minimal model, the correction to $Z \bar{b}_L b_L$ coupling is given by 
\begin{align}
\delta g_L^{bb}  \sim 1.5 \times 10^{-2}  (\eps^Q_3)^2 (\lambda_*^D)^2 \left( \frac{1 \, \TeV}{M_D} \right)^2, 
\end{align}
which is $\sim 9 \times 10^{-4}$ for the above value of $M_D$ implying that $|\delta g_L^{bb}/g_{L,SM}^{bb}| \sim 2 \times 10^{-3}$ and there is no conflict with electroweak precision data~\cite{ALEPH}. This is a consequence of the general relation $\delta g_L^{ds}/\delta g_L^{bb} \sim \eps^5$ as both couplings have the same parametric dependence up to the flavor dependence of $\lambda^D_{ij}$. One therefore expects that for flavor anarchic $\lambda^D_{ij}$ the bound on $\delta g_L^{ds}$ is as (or even more) important than the one on $\delta g_L^{bb}$ since $\delta g_L^{ds} \lesssim 6 \times 10^{-7}$ implies $|\delta g_L^{bb}/g_{L,SM}^{bb}| \lesssim 2 \times 10^{-3}$. Note however that the bounds from $K_L \rightarrow \mu^+ \mu^-$ involve certain non-perturbative uncertainties. Other bounds arise from the contribution of heavy fermions to the electroweak precision observables $S$ and $T$ which are roughly comparable to the flavor physics bounds~\cite{Agashe:2003zs, Simplified, AgSo}. We will provide an extensive discussion of the various phenomenological aspects in much greater detail in Ref.~\cite{BGPZ2}.

It is remarkable that already the minimal FCNC effects that we are studying here give quite stringent bounds at least on $M_D$, not much weaker than the typical constraints on the compositeness/KK scale one finds in complete models \cite{BDG,RSbounds}. On the other hand, the bounds on $M_{Q,U,D}$ do not exclude the possibility that some of the heavy fermions could be as light as a TeV.

Finally we comment on the impact of departing from the minimal model by including $\gamma \ne 0$ with typical strength $\gamma_*$. In this case Higgs mediated FCNCs are no longer negligible, but the bounds on heavy fermions increase only slightly. For example the coefficient of $(\bar s_R\, d_L)(\bar s_L d_R)/( \TeV)^2$ can be estimated to be 
\begin{align}
C_{LR} \approx 50 \left(\frac{200 \, \GeV}{m_H}\right)^2 \frac{v^2 m_s m_d}{M_D^2 M_Q^2} \gamma_*^2 (\lambda^D_*)^2,
\end{align}
and the bounds from $\eps_K$ can be satisfied by taking e.g. $M_D \gtrsim  4 \, \TeV \times \gamma_* \lambda_*^D$ and $M_Q \gtrsim  1.2 \, \TeV \times \gamma_* \lambda_*^D $ for a Higgs mass $m_H = 115$ GeV.

\section{Connection to Existing Models}
\label{AS}
Finally we want to outline the connection of our model with existing flavor models by comparing complete flavor models with the general Lagrangian (\ref{Lgen})
\begin{align}
{\cal L} \sim m f \Fbar + M F \Fbar + \lambda h F F + \gamma h \Fbar \, \Fbar + \alpha h f F.    
\end{align}
The bounds on heavy fermion masses derived in Section \ref{pheno} from $Z$ mediated FCNCs should apply to these models as well. Of course in complete models there are typically additional (and stronger) sources of FCNCs which however depend on the specific model under consideration.  

Lagrangian (\ref{Lgen}) has been widely used as a simplified formulation~\cite{Simplified,CoPo} for the fermionic sector of composite Higgs models~\cite{CompHiggs, Giudice:2007fh} that are dual to certain warped 5d models~\cite{Csaki:2003zu, compduals}. In the first picture the chiral fermions represent weakly coupled elementary fields, while the vectorlike fermions are composite states that are part of a strongly interacting sector with typical masses around the TeV scale. The composite sector includes the Higgs boson which does not couple to elementary states implying $\alpha = 0$. The SM fermions are then linear combinations of elementary and composite fields, typically with small components in the composites. Their couplings to the Higgs are then suppressed by these small mixing angles that correspond to our $\eps$'s which in this context can be interpreted as the amount of compositeness. To see the relation to warped 5d models one has to diagonalize the explicit mass terms in Lagrangian (\ref{Lgen}). The massive fermions correspond to the higher KK states, while three chiral fermions remain massless and represent the KK zero-modes. They receive their mass from EWSB through their couplings to the Higgs that is localized on the IR brane. These zero-modes have a wave function profile that is governed by the 5d bulk masses. This profile has by construction an exponentially hierarchical structure which suppresses all couplings of the zero-modes and corresponds to our $\eps$'s. In both pictures the couplings of the Higgs to heavy fermions are usually taken to be anarchical and $\ord{1}$.

Let us now compare our results for heavy fermions with the studies that have been done in the context of composite higgs models \cite{Simplified, AgCo}. In Ref.~\cite{AgCo}  the bounds on heavy fermion masses have been obtained from tree-level higgs exchange contributions to the FCNC processes, for  flavour violating couplings to the higgs boson corresponding to our general case with
$\gamma \ne 0$.  It was shown that the constraints on the compositeness scale (identified with the vector-like fermion mass scale) are generically as strong as from the exchange of heavy spin-1 resonances. In Refs.~\cite{Giudice:2007fh, AgCo} it is also pointed out that our special case of $\gamma$ aligned with the SM Yukawa couplings (effectively similar to  $\gamma=0$) corresponds to a composite pseudo-Goldstone Higgs boson and implies much milder bounds, in agreement with our results. In Ref.~\cite{Simplified} a similar approach is used to discuss the bounds on heavy fermion masses and on spin 1  resonances from precision electroweak data ($Z \bar{b}b$ coupling, $S$ and $T$).  One of the main results of our paper is that the most generic implication of the effective theory of fermion masses based on the Lagrangian (1) are modifications of the $Z$ boson couplings that lead to the bounds on the heavy fermion masses as strong (or even stronger) than the previously discussed bounds.

Our second point is that the effective model based on Lagrangian (1) captures not only the flavour physics of composite higgs models but also that of Froggatt--Nielsen models \cite{HSymm}. Since this point has been discussed less often, we give  below some details. In this case the heavy fermions play the role of messengers that communicate flavor symmetry breaking to the light fields. The mass terms $m$ mixing light fields and messengers arise from flavor symmetry breaking and are parameterized by the VEV of a single scalar field $\vev{\phi}$ in the simplest cases. Typically a large number of messengers is introduced and by successively integrating out\footnote{Instead of integrating out, one can think of decoupling them after diagonalizing their mass terms.} these heavy fermions higher powers of $\vev{\phi}/M$ are generated in the Lagrangian. Let us start by writing the schematic Lagrangian for the down-type sector and one flavon $\phi$ 
\begin{align}
\label{Lstart}
{\cal L} & \sim  M_Q \overline{Q} Q + M_D \overline{D^c} D^c + a_Q \phi \overline{Q} Q + a_D \phi \overline{D^c} D^c  + b_Q \phi \overline{Q} \, q + b_D \phi \overline{D^c} \, d^c \nonumber \\ 
& + \lambda h Q \, D^c + \gamma h \overline{Q} \, \overline{D^c} + \alpha h Q \, d^c + \beta h D^c \, q, 
\end{align}
where $a, b, \lambda, \gamma, \alpha, \beta$ are $\ord{1}$ matrices whose structure is determined by the flavor symmetry. Obviously this Lagrangian is of the form (\ref{Lgen}) once we replace the flavon by its VEV. In order to make contact to our minimal model we have to integrate out all messengers but three in each sector. For each messenger that we integrate out we can get a small factor $\vev{\phi}/M$ entering the effective couplings. To ensure that these factors suppress only the effective masses $m$ and not the effective couplings $\lambda$, we want to integrate out only those messengers which do not have $\lambda$ couplings. Moreover we need to require that no messengers have Yukawa couplings with light fields.  To recover the structure of the minimal model (with $\gamma \ne 0$) we therefore impose\footnote{Such a structure could be enforced by the horizontal symmetry, and below we will give an example for a simple $U(1)$ model.} that 1) $\alpha = \beta = 0$, 2) $M_Q$ and $M_D$ approximately universal and 3) rank $\lambda = 3$, such that only three $D$'s (and $U$'s) couple to (the same) three $Q$'s. If we now integrate out the messengers that do not have Yukawa couplings by their EOMs which are of the schematic form
\begin{align}
Q & \sim  a_Q \frac{\phi}{M_Q} Q + b_Q \frac{\phi}{M_Q} q + \gamma \frac{h}{M_Q} \overline{D^c} & \overline{Q} & \sim a_Q \frac{\phi}{M_Q} \overline{Q} \\ 
D^c & \sim a_D \frac{\phi}{M_D} D^c + b_D \frac{\phi}{M_D} d^c + \gamma \frac{h}{M_D} \overline{Q} & \overline{D^c} & \sim a_D \frac{\phi}{M_D} \overline{D^c},  
\end{align}
we obtain an effective Lagrangian that is again of the form (\ref{Lstart}), but with certain powers of the suppression factor $\vev{\phi}/M$ entering the effective couplings $a_{Q,D}, b_{Q,D}, \gamma$. Therefore the mere effect of integrating out heavy fields is to generate small mass terms $m$ and small couplings $\gamma$ while $\lambda$ remains $\ord{1}$. This means that the fermionic sector of any flavor model which satisfies the above conditions is effectively described by the minimal model (\ref{Lmin}), generalized to include $\gamma \ne 0$. 

In order to illustrate that these conditions are not very difficult to satisfy, we consider the UV completion of a simple $U(1)$ FN model that reproduces the light fermion mass matrices we used in Section \ref{pheno}. To this end we choose $U(1)$ charges of the light fields as $u^c_{1,2,3} = (5,2,0)$, $d^c_{1,2,3} = (5,4,3)$ and $q_{1,2,3} = (3,2,0)$ and introduce a single flavon $\phi$ with charge $-1$, and take the Higgs to be uncharged. We then introduce FN messengers $U^c_{4,3,2,1,0},D^c_{4,3,2,1,0},Q_{2,1,0}$ plus conjugates which we labeled by their $U(1)$ charge. In addition we have to add a certain number of these messenger fields with the same $U(1)$ quantum numbers in order to reproduce SM masses \cite{HSymm}, so that the total number of messengers with $u^c$ ($d^c$) SM quantum numbers is 12 (17). So we can take in total three copies of $U^c_0,D^c_0,Q_0$ and say 6 copies of $D_1$ and still have only the minimal number of messengers needed. If we (as usual) require that the messenger masses are approximately degenerate in each sector, it is obvious that our conditions are satisfied, since apart from the third family (where the mixing is large and $\beta$ can be absorbed into $\lambda$) only the uncharged messengers have Yukawa couplings.

\section{Additional Structure: Unitary Model}
Finally one can take a bottom-up approach and consider Eq.~(\ref{L3}) as a minimal extension of the SM that parameterize small Yukawa couplings through mass hierarchies. In order to make this model more predictive, one can try to reduce the number of free parameters, for example by imposing additional structure in heavy fermion sector. One interesting possibility is to impose a flavor symmetry that is intact in the heavy sector and is broken only by mass terms $m^{Q,U,D}$. This could explain the absence of light-heavy Higgs couplings and would force $M^{Q,U,D}$ and $\lambda^{U,D}$ to be universal and unitary respectively, provided that heavy fermions transform in three-dimensional representations under the flavor group. 

While a unitary SM Yukawa matrix is clearly ruled out by the data, a unitary Yukawa matrix $\lambda$ in the heavy fermion sector is not only allowed phenomenologically but turns out to be rather appealing for the following reasons:
\begin{itemize}
\item
The number of fundamental parameters is reduced to 18 real and 4 phases, implying only 9 new real parameters and three new phases. This may allow to obtain interesting correlations of flavor observables. 
\item
As seen in Eq.~(\ref{ALXARX}) in the case of a unitary $\lambda$ the corrections to the SM $Z$ couplings to light fermions become diagonal in the flavor space with the diagonal entries being proportional to $\epsilon_i^2$. After the rotation to the mass eigenstates the non-universality of $\epsilon_i$, necessary for the explanation of fermion mass hierarchies, induce tree level FCNC transitions which are suppressed not only by heavy fermion masses $M^X$ but in the case of two first generations by the differences in light quark masses.
\item
Similarly as seen in Eq.~(\ref{effy}) also the corrections to the SM Higgs-fermion  couplings become diagonal in the flavor basis implying also suppression of Higgs mediated FCNC processes.
\end{itemize}
We will analyze the viability of this scenario in more detail in Ref.~\cite{BGPZ2}. 

\section{Conclusions}
In this paper we have considered a minimal theory of fermion masses, obtained from extending the SM by heavy vectorlike fermions that mix with chiral fermions such that small SM Yukawa couplings arise from small mixing angles. Since under certain conditions this model can serve as an effective description of the fermionic sector of a large class of existing flavor models, it might be regarded as a useful reference frame for a further understanding of flavor hierarchies in the SM. We emphasized that already such a minimal framework gives rise to FCNC effects through exchange of massive SM bosons whose couplings to the light fermions get modified by the mixing. We derived these couplings and used the results to put lower bounds on the masses of the heavy fermions. Particularly stringent bounds, in a few TeV range, come from the corrections to the $Z$ couplings. Still, they do not exclude the possibility that the mass scale of the heavy fermions could be related to certain dynamics responsible for stabilizing the weak scale. Some of the heavy fermions can also be lighter than a TeV and thus might be in the reach of the LHC. We outlined the connection of the minimal theory with complete flavor models such as FN and RS models and discussed additional structures that could be imposed on the heavy fermionic sector. Particularly appealing seems to be the possibility that heavy Yukawa couplings are unitary matrices, a situation that might be enforced by flavor symmetries that remain unbroken in these sectors. An extensive analysis of several aspects related to the ideas presented in this paper will be performed in Ref.~\cite{BGPZ2}.

\section*{Acknowledgments}
S.P. acknowledges support by the MNiSZW grant N N202 103838(2010-2012). This work has been partially done in the context of the ERC Advanced Grant ``FLAVOUR'' (267104) and has been partly supported by the European Commission under the contract ERC Advanced Grant 226371 MassTeV and the contract PITN-GA-2009-237920 UNILHC. R.Z. would like to thank the CERN Theory Division for kind hospitality during the completion of this work.

\end{document}